\documentclass[floatfix,aps,prl,twocolumn,superscriptaddress]{revtex4}
\usepackage[dvipdfmx]{graphicx}
\usepackage[]{epsfig}
\usepackage{times,amsmath,amssymb}
\begin{document}

\title{Four single-spin Rabi oscillations in a quadruple quantum dot}

\author{Takumi Ito}
\email[]{takumi.ito@riken.jp}
\affiliation{Center for Emergent Matter Science, RIKEN, 2-1 Hirosawa, Wako, Saitama 351-0198, Japan}
\affiliation{Department of Applied Physics, University of Tokyo, Bunkyo, Tokyo 113-8656, Japan}

\author{$^{,\dagger} $Tomohiro Otsuka}
\altaffiliation{\ These authors contributed equally to this work.} 
\affiliation{Center for Emergent Matter Science, RIKEN, 2-1 Hirosawa, Wako, Saitama 351-0198, Japan}
\affiliation{Department of Applied Physics, University of Tokyo, Bunkyo, Tokyo 113-8656, Japan}
\affiliation{JST, PRESTO, 4-1-8 Honcho, Kawaguchi, Saitama, 332-0012, Japan}

\author{Takashi Nakajima}
\affiliation{Center for Emergent Matter Science, RIKEN, 2-1 Hirosawa, Wako, Saitama 351-0198, Japan}
\affiliation{Department of Applied Physics, University of Tokyo, Bunkyo, Tokyo 113-8656, Japan}

\author{Matthieu R. Delbecq}
\affiliation{Center for Emergent Matter Science, RIKEN, 2-1 Hirosawa, Wako, Saitama 351-0198, Japan}
\affiliation{Department of Applied Physics, University of Tokyo, Bunkyo, Tokyo 113-8656, Japan}

\author{Shinichi Amaha}
\affiliation{Center for Emergent Matter Science, RIKEN, 2-1 Hirosawa, Wako, Saitama 351-0198, Japan}

\author{Jun Yoneda}
\affiliation{Center for Emergent Matter Science, RIKEN, 2-1 Hirosawa, Wako, Saitama 351-0198, Japan}
\affiliation{Department of Applied Physics, University of Tokyo, Bunkyo, Tokyo 113-8656, Japan}

\author{Kenta Takeda}
\affiliation{Center for Emergent Matter Science, RIKEN, 2-1 Hirosawa, Wako, Saitama 351-0198, Japan}
\affiliation{Department of Applied Physics, University of Tokyo, Bunkyo, Tokyo 113-8656, Japan}

\author{Akito Noiri}
\affiliation{Center for Emergent Matter Science, RIKEN, 2-1 Hirosawa, Wako, Saitama 351-0198, Japan}
\affiliation{Department of Applied Physics, University of Tokyo, Bunkyo, Tokyo 113-8656, Japan}

\author{Giles Allison}
\affiliation{Center for Emergent Matter Science, RIKEN, 2-1 Hirosawa, Wako, Saitama 351-0198, Japan}

\author{Arne Ludwig}
\affiliation{Angewandte Festk\"{o}rperphysik, Ruhr-Universit\"{a}t Bochum, D-44780 Bochum, Germany}

\author{Andreas D. Wieck}
\affiliation{Angewandte Festk\"{o}rperphysik, Ruhr-Universit\"{a}t Bochum, D-44780 Bochum, Germany}

\author{Seigo Tarucha}%
\email[]{tarucha@ap.t.u-tokyo.ac.jp}
\affiliation{Center for Emergent Matter Science, RIKEN, 2-1 Hirosawa, Wako, Saitama 351-0198, Japan}
\affiliation{Department of Applied Physics, University of Tokyo, Bunkyo, Tokyo 113-8656, Japan}
\affiliation{Quantum-Phase Electronics Center, University of Tokyo, Bunkyo, Tokyo 113-8656, Japan}
\affiliation{Institute for Nano Quantum Information Electronics, University of Tokyo, 4-6-1 Komaba, Meguro, Tokyo 153-8505, Japan}

\date{\today}
\begin{abstract}
Scaling up qubits is a necessary step to realize useful systems of quantum computation.
Here we demonstrate coherent manipulations of four individual electron spins using a micro-magnet method in a quadruple quantum dot - the largest number of dots used for the single spin control in multiple quantum dots.
We observe Rabi oscillations and electron spin resonance (ESR) for each dot and evaluate the spin-electric coupling of the four dots, and finally discuss practical approaches to independently address single spin control in multiple quantum dot systems containing even more quantum dots.
\end{abstract}

\maketitle

Semiconductor quantum dots (QDs) provide a promising platform for quantum information processing in solid state devices~\cite{1995DiVicenzoScience,1998LossPRA,2005PettaScience,2006KoppensNature}.
The QD system has advantages in implementing quantum bits (qubits), including a relatively long coherence time of electron spin, potential scalability thanks to the well-established fabrication technology and small physical size per qubit.
To date gate-defined QDs~\cite{2000CiorgaPRB} have been up-scaled to double~\cite{1996BlickPRB}, triple~\cite{2006GaudreauPRL}, quadruple~\cite{2014TakakuraApplPhysLett,2014DelbecqApplPhysLett} and quintuple QDs~\cite{2016ItoSciRep} to increase the number of qubits.
Recently half-filled QDs were demonstrated in an array of nine QDs~\cite{2016ZajacPRA}.
The multiple QDs can also be used to study the physics of electron-electron interactions such as quantum cellular automata~\cite{2006GaudreauPRL} and the Fermi-Hubbard model~\cite{2017HensgensNature}.
In parallel with up-scaling of the QD system, basic manipulations of a few qubits have been demonstrated~\cite{2005PettaScience,2007NowackScience,2008PioroNatPhys,2011BrunnerPhysRevLett}.
We used a micro-magnet ESR (MM-ESR) method to realize the fastest control of a single spin~\cite{2014YonedaPhysRevLett}, and then individual control of electron spins confined in a triple QD~\cite{2016NoiriAPL} and four individual ESR signals in a quadruple QD (QQD)~\cite{2016OtsukaSciRep}.

\begin{figure}
\begin{center}
  \includegraphics{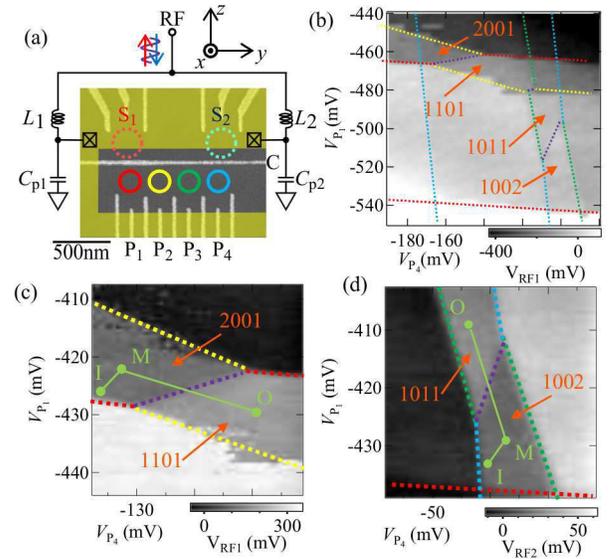}
  \caption{(color online)
(a) Scanning electron microscope image of the QQD device with a schematic representation of the measurement electric circuit.
(b) $V_{\rm RF1}$ of $V_{\rm P_1}$ and $V_{\rm P_2}$ (stability diagram) with a background plane subtracted.
At the lower left corner of this diagram, all dots are completely depleted ([0,0,0,0] charge state).
(c) Enlarged stability diagram relevant for experiments in QD$_1$ and QD$_2$.
Green circles and bars show the voltage conditions and pulse shape that are utilized in ESR measurements. 
(d) Enlarged stability diagram relevant for experiments in QD$_3$ and QD$_4$.
}
\label{fig1}
\end{center}
\end{figure}

In this paper, we perform single spin control in a QQD with the MM-ESR method.
We use a correlated double sampling (CDS) technique~\cite{2014ShulmanNatCom} to enhance the ESR signal as compared to our previous experiment~\cite{2016OtsukaSciRep} and observe a coherent oscillation (Rabi oscillation) of an electron spin in each dot by sweeping the ESR driving time.
Based on the acquired data, we discuss how to improve the quality of the Rabi oscillations by considering the control speed and the addressability of the electron spin in each dot.

Fig.\ref{fig1} (a) shows a scanning electron micrograph of the QQD device and a schematic picture of the measurement electric circuit.
The device is fabricated in a GaAs/AlGaAs modulation doped two dimentional electron gas containing heterostructure wafer.
The Ti/Au gate electrodes placed on the surface appear in white.
The MM is placed on top of the gate electrode layer with a 100 nm thick insulator in between (shown in yellow shaded region).
By applying negative voltages on the gate electrodes, we form six dots in total as pictorially indicated by four solid circles and two dotted circles.
The four dots under the horizontal gate electrode named C are the qubit dots named QD$_1$, QD$_2$, QD$_3$ and QD$_4$ in red, yellow, green and blue, respectively and tunnel-coupled next to each other.
The two dots above gate C are the sensor dots named S$_1$, and S$_2$ in red, and blue, respectively.
The sensor dots are connected to the respective RF resonators configured by inductances  $L_1$ and $L_2$ and stray capacitances $C_{\rm p1}$ and $C_{\rm p2}$ with resonance frequency of $f_{\rm RES1}$=298 MHz for S$_1$ and $f_{\rm RES2}$=207 MHz for S$_2$.
We monitor the charge state of the QQD via the reflected RF signals~\cite{2007ReillyAPL,2010BarthelPRB}, $V_{\rm RF1}$ and $V_{\rm RF2}$ of the two sensors at the respective resonance frequencies.
The MM-ESR is performed by applying a microwave (MW) to gate C in the presence of an external magnetic field $B_{\rm ext}$ along the z-axis as shown in Fig.\ref{fig1} (a).
The MM magnetizes in the $B_{\rm ext}$ direction and creates a stray field across the QQD.
The shape of the MM is specially designed for the MM-ESR to address the four dots~\cite{2016NoiriAPL,2016OtsukaSciRep,2015YonedaAPEX} such that the stray field produces a slanting field $B_{\rm Sl}$ along x with z for driving the electron spin rotation and a local magnetic field $B_{\rm Z}$ along z that changes the resonance condition among the four dots.
All measurements described below are conducted for the QQD device placed in a dilution fridge at a temperature of $T=13$ mK.

Fig.\ref{fig1} (b) shows the charge stability diagram measured by monitoring  $V_{\rm RF1}$ as a function of the voltages $V_{\rm P_1}$ and $V_{\rm P_4}$ of the plunger gates P$_1$ and P$_4$, respectively.
We identify the dot-lead and inter-dot charge transition lines indicated by the dotted lines, and assign them in red, yellow, green and blue from horizontal to vertical to the dot-lead charge transition lines of QD$_1$, QD$_2$, QD$_3$ and QD$_4$, respectively.
The inter-dot charge transition lines are observed as the lines in purple between the opposite cross points of two dot-lead charge transition lines.
We denote the charge state as [$n_1$,$n_2$,$n_3$,$n_4$], where $n_i$ with $i$= 1 to 4 is denoting the number of electrons confined in QD$_i$.
In the MM-ESR experiment we use Pauli spin blockade (PSB) in tunnel-coupled  double QDs (DQDs) for the spin readout~\cite{2005PettaScience}.
Then the spin configuration, up or down is distinguished by measuring the transition between [2,0] (or [0,2]) and [1,1].
So in the following measurement we separate the QQD into two DQDs of QD$_1$ - QD$_2$ and QD$_3$ - QD$_4$, and mainly focus on the charge states of [2,0,0,1] and [1,1,0,1], and [1,0,1,1] and [1,0,0,2] to operate the MM-ESR in QD$_1$ and QD$_2$ with S$_1$, and QD$_3$ and QD$_4$ with S$_2$, respectively.
Figs.\ref{fig1} (c) and (d) show the enlarged stability diagrams around the boundary line between [2,0,0,1] - [1,1,0,1], and [1,0,1,1] - [1,0,0,2], respectively. 

\begin{figure}
\begin{center}
  \includegraphics{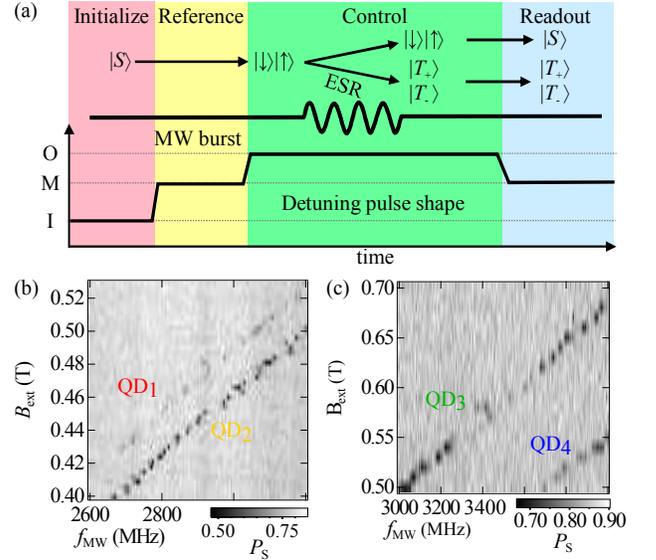}
  \caption{(color online)
(a) The pulse sequence for the ESR measurement.
The horizontal axis shows time and the vertical axis pictorially denotes the voltage conditions at points O, M and I in Figs. \ref{fig1} (c) and (d).
Measured $P_{\rm S}$ for QD$_1$ and QD$_2$ (b) and QD$_3$ and QD$_4$ (c) in the plane of $B_{\rm ext}$ and $f_{\rm MW}$ with application of MW.
The two black lines in each figure indicate the ESR lines for two different QDs.   
}
\label{fig2}
\end{center}
\end{figure}

Fig.\ref{fig2} (a) shows a schematic of the gate voltage pulses for the ESR measurements.
Here three voltage conditions named I, M and O are specified in Figs.\ref{fig1} (c) and (d) by the green circles to define the four operation sections, ``Initialization" in red, ``Reference" in yellow, ``Control" in green and ``Readout" in blue.
In the ``Initialization" section, the voltage condition is tuned to point I that is close to the charge transition line of the outer dots (QD$_1$ and QD$_4$).
The two-electron spin state in QD$_1$ - QD$_2$ (or QD$_3$ - QD$_4$) is initialized to the ground doubly occupied singlet state $|S\rangle$ in QD$_1$ (or QD$_4$) by exchanging electrons with the adjacent reservoir.
In the ``Reference" section the voltage condition is tuned to point M that is deep inside the Coulomb blockade region of the doubly occupied charge state [2,0,0,1] (or [1,0,0,2]).
This stage does not influence the spin state and we obtain the background signal of the charge sensor in this section.
In the ``Control" section the voltage condition is ramped to point O where two electrons in QD$_1$ (or QD$_4$) are separated into QD$_1$ and QD$_2$ (or QD$_3$ and QD$_4$) to form [1,1,0,1] (or [1,0,1,1]) having the anti-parallel spin state $\mid\downarrow\rangle \mid\uparrow\rangle$~\cite{2005PettaScience, 2016OtsukaSciRep}.
Then, a MW burst is applied on gate C to spatially oscillate electron spins in all dots in the MM induced $B_{\rm Sl}$.
The electron spin coherently flips between $\mid\uparrow\rangle$ and $\mid\downarrow\rangle$ but independently in each QD when the resonance condition of $f_{\rm MW} =|g|\mu_{\rm B}(B_{\rm ext}+B_{\rm z})/h$ is satisfied.
Here $f_{\rm MW} $ is the MW frequency, $g$ is the Lande $g$-factor of the electron confined in the QD, and $\mu_{\rm B}$ is the Bohr magneton, respectively.
Finally, in the ``Readout" section with the voltage condition back to point M, the spin state is detected using PSB.
When the spin is not flipped in either of QD$_1$ or QD$_2$ (or QD$_3$ or QD$_4$), the two-electron spin state returns to the doubly occupied $|S\rangle$ in QD$_1$ (or QD$_4$).
When the spin is flipped in either dot, the two-electron state is either  $|T_{\rm +}\rangle$ or $|T_{\rm -}\rangle$, the charge state remains singly occupied in QD$_1$ - QD$_2$ (or QD$_3$ - QD$_4$) without relaxing to the doubly occupied $|S\rangle$.
Here, we measure the singlet return probability $P_{\rm S}$ or probability of finding the two electron spin state in the $|S\rangle$ by taking the difference of $V_{\rm RF1}$ (or $V_{\rm RF2}$) from that measured in the "Reference" section (CDS method).
Note this CDS technique is efficient to compensate the low frequency noise in the charge sensor.

\begin{figure}
\begin{center}
  \includegraphics{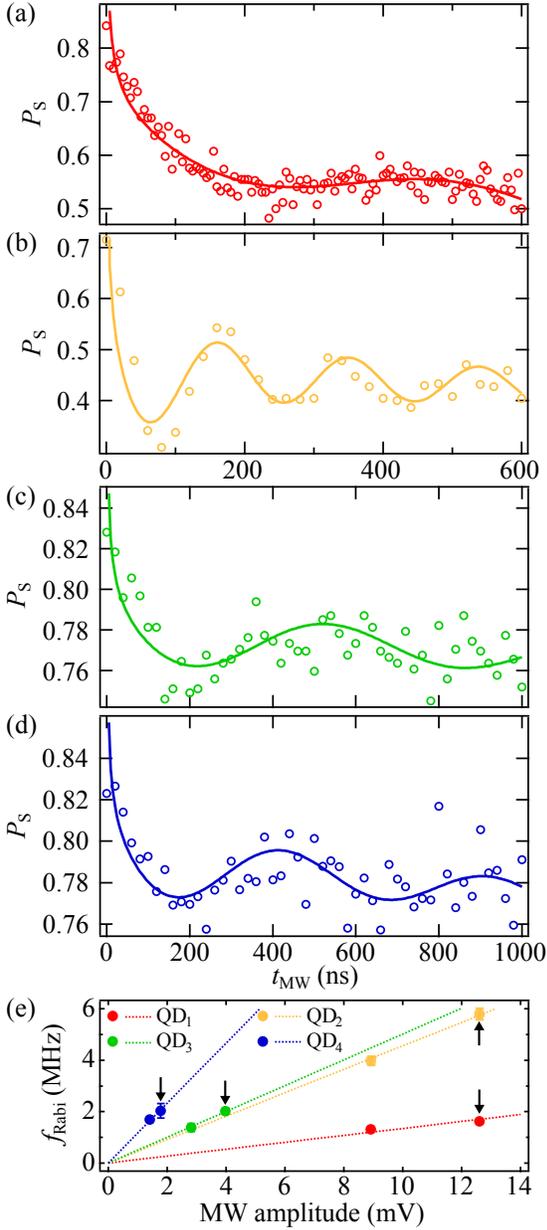}
  \caption{(color online)
Rabi oscillations of the single electron spins measured for 
QD$_1$ with $f_{\rm MW}=2751$ MHz, $B_{\rm ext}=0.45$ T and $P_{\rm MW}=10$ dBm in (a),
QD$_2$ with $f_{\rm MW}=2280$ MHz, $B_{\rm ext}=0.45$ T and $P_{\rm MW}=10$ dBm in (b),
QD$_3$ with $f_{\rm MW}=3650$ MHz, $B_{\rm ext}=0.6$ T and $P_{\rm MW}=0$ dBm in (c),
and QD$_4$ with $f_{\rm MW}=4400$ MHz, $B_{\rm ext}=0.6$ T and $P_{\rm MW}=-6$ dBm in (d).   
The fitting parameters are
$A=0.73 \sqrt{\rm ns}, f_{\rm Rabi}=1.62$ MHz, $B_{\rm off}=0.65, C=0.26 \mu$s$^{-1}$ for QD$_1$, 
$A=0.81 \sqrt{\rm ns}, f_{\rm Rabi}=5.32$ MHz, $B_{\rm off}=0.46, C=0.051 \mu$s$^{-1}$ for QD$_2$,
$A=0.38 \sqrt{\rm ns}, f_{\rm Rabi}=2.02$ MHz, $B_{\rm off}=0.69, C=0.012 \mu$s$^{-1}$ for QD$_3$,
$A=0.22 \sqrt{\rm ns}, f_{\rm Rabi}=2.03$ MHz, $B_{\rm off}=0.79$ and $C=0.018 \mu$s$^{-1}$ for QD$_4$.
(e) MW amplitude dependence of $f_{\rm Rabi}$ derived for each dot from the curve fitting to the Rabi oscillation data.
Data points extracted from (a) to (d) are indicated by black arrows.
The dotted lines are the fitting to the data points including the origin. 
}
\label{fig3}
\end{center}
\end{figure}

Figs.\ref{fig2} (b) and (c) show $P_{\rm S}$ as a function of $f_{\rm MW}$ and $B_{\rm ext}$ upon application of the MW pulse measured using S$_1$ and S$_2$, respectively.
We observe two distinct lines due to ESR in QD$_1$ and QD$_2$ in (b) and QD$_3$ and QD$_4$ in (c).
The separation of the two ESR lines in each figure is due to the Zeeman energy difference, which arises from the differences in the MM-induced $B_{\rm Z}$ and $g$-factor among dots.
This separation is much larger than the ESR line width of $\sim10$ MHz and therefore enables us to access each resonance condition independently by choosing the $f_{\rm MW}$ and $B_{\rm ext}$ properly.


Next, we perform measurements of single-electron spin oscillations.
We apply the same voltage pulse sequence as used in the ESR measurement but change the MW burst time $t_{\rm MW}$ in the ``Control" section to see Rabi oscillations.
Figs.\ref{fig3} (a), (b), (c), and (d) show the $P_{\rm S}$ measured at the respective resonance condition of QD$_1$, QD$_2$, QD$_3$ and QD$_4$.
We observe Rabi oscillations of the electron spin in each dot.
The Rabi oscillations are best resolved in QD$_2$ with frequency 5.32 MHz, the highest among all dots.
On the other hand, the Rabi oscillation is less clear for the other QDs because of the lower frequency of about 2 MHz.
Nevertheless, all Rabi oscillation data are well fitted by a power law envelope function with a $\pi/4$ phase shift~\cite{2007KoppensPRL} $P_{\rm S}=\frac{A}{\sqrt{t_{\rm MW}}}$cos$(2\pi f_{\rm Rabi}t_{\rm MW}+\pi/4)+B_{\rm off}-Ct_{\rm MW}$.
The last linear term accounts for the reduction of $P_{\rm S}$ due to the leakage to non-qubit states presumably caused by photon-assisted tunneling~\cite{2010NadjaNature}.

\begin{figure}
\begin{center}
  \includegraphics{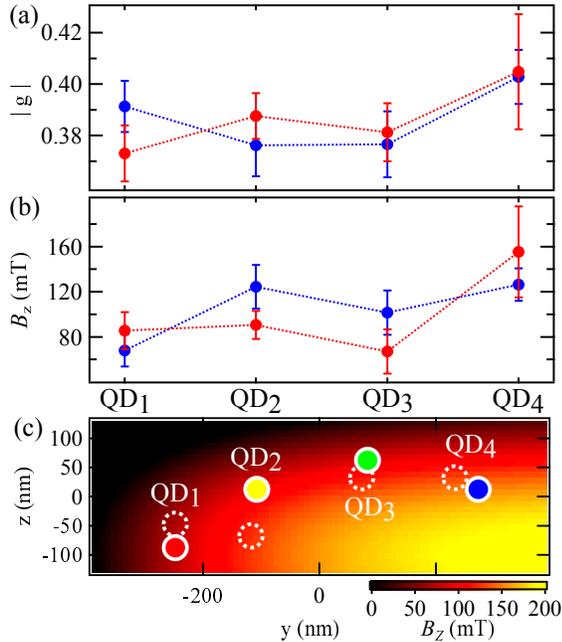}
  \caption{(color online)
(a) $g$ factors in each dot estimated from Figs.\ref{fig2} (d) and (e) (colored in red).
(b) $B_{\rm Z}$ in each dot estimated from the same data with (b) (colored in red).
For comparison, $g$ factors and $B_{\rm Z}$ values estimated from the data in a different gate voltage condition~\cite{2016OtsukaSciRep} are also shown as blue circles in (a) and (b).
(c) Spatial distribution of the numerically calculated $B_{\rm Z}$ and the dot positions construed from the result in (b).
The solid, and dotted circles indicate the positions for the data set in red, and blue in (b), respectively.
Here we define the origin of the y and z axis as the center of the QD array.
}
\label{fig4}
\end{center}
\end{figure}

We measured Rabi oscillations for various MW output power $P_{\rm MW}$ values and derived $f_{\rm Rabi}$ from the curve fitting described above.
The obtained values of $f_{\rm Rabi}$ are shown as a function of MW amplitude in Fig.\ref{fig3} (e).
The MW amplitude is calculated from the applied $P_{\rm MW}$ and the RF line attenuation solely given by discrete attenuators ($-39$ dB).
We observe $f_{\rm Rabi}$ linearly depending on the MW amplitude for each QD.
We derive the slope of the linear fitting to the data points and find that it is different from dot to dot with ratio of $3:10:11:26$ for QD$_1$ to QD$_4$.
The slope of the fitting line is related to the spin-electric coupling roughly proportional to the product of $B_{\rm Sl}$ and $l_{\rm orb}^2/\Delta$ where $l_{\rm orb}$ is the orbital spread along z axis and $\Delta$ is the QD confinement energy.
Considering the MM design used in this device, the $B_{\rm Sl}$ value should gradually increase from QD$_1$ to QD$_4$.
The experimental data seemingly agree with this trend but the observed variation of the slope is quantitatively larger than expected.
We discuss this discrepancy later using Fig.\ref{fig4}.

We note that we could not apply a large $P_{\rm MW}$ to the right DQD (QD$_3$ and QD$_4$) while we could to the left DQD (QD$_1$ and QD$_2$), because the ESR signals of QD$_3$ and QD$_4$ become obscure in the high $P_{\rm MW}$ range.
So the Rabi oscillations shown in Figs. \ref{fig3} (c) and (d) are only measured at a small $P_{\rm MW}$, and therefore the oscillation frequency is significantly lower than that for QD$_2$.
This may not be related to the robustness of the ESR conditions because $P_{\rm S}$ decreases with increasing $P_{\rm MW}$ even in the off-resonance conditions.
One of the possible reasons is that the tunnel barrier between QD$_4$ and the right reservoir is not well closed and the electron can tunnel out to the reservoir more easily as the $P_{\rm MW}$ becomes large.


In what follows, we discuss the electron spin addressability in the QQD device.
We estimate the $g$-factors and $B_{\rm Z}$ values of individual QDs from the resonance lines in Figs.\ref{fig2} (b) and (c).
These values are shown by the red circles in Figs. \ref{fig4} (a) and (b).
Those extracted from the data measured with the different gate voltage condition in our previous experiment~\cite{2016OtsukaSciRep} are also shown by the blue circles.
With both variations of $g$ and $B_{\rm Z}$ we are able to independently address  the MM-ESR in each dot.
Indeed, the $B_{\rm Z}$ difference alone will not be large enough to resolve resonances between QD$_1$ and QD$_2$ in the present experiment, because the ESR line separation will be only 20 MHz comparable to the ESR line width.
The variation of $g$ between QDs may be explained by the difference in the confinement potential~\cite{2018MichalPRB}.
On the other hand, that of $B_{\rm Z}$ is probably due to variation of the dot position and inhomogeneity of the MM induced stray field.
In support of this, we find the observed $B_{\rm Z}$ in each dot different when changing the gate voltage conditions.
This result implies that $B_{\rm Z}$ is varied from dot to dot in a controlled manner with the gate voltage condition and therefore allows us to independently address the MM-ESR in each dot.
Fig.\ref{fig4} (c) shows a two-dimensional distribution (y-z plane in Fig.\ref{fig1} (a)) of the $B_{\rm Z}$ calculated numerically from the shape of the MM and the QD positions that explain the results of Fig.\ref{fig4} (b).
The solid, and dotted circles indicate the positions for the data set in red, and blue in Fig.\ref{fig4} (b), respectively.
The variation of the dot positions suggests that the QDs are formed in a disordered manner due to e.g. charge impurities.

Finally, we discuss the variation of the MW amplitude dependence of $f_{\rm Rabi}$ among the four dots as observed in Fig.\ref{fig3} (e).
The dot positions which can explain the $B_{\rm Z}$ values shown in Fig.\ref{fig4} (c) give the $B_{\rm Sl}$ ratio of $3:3:2:4$ for QD$_1$ to QD$_4$.
These $B_{\rm Sl}$ values are numerically calculated in the same way with $B_{\rm Z}$.
This variation of $B_{\rm Sl}$ among dots is too small to fully account for that of the slopes of MW amplitude dependence.
The variation of the $g$ as shown in Fig.\ref{fig4} (a) can also influence the control speed but not so significantly.
This discrepancy may be caused by inhomogeneity of the spin-electric coupling, which depends on the inhomogeneity of the confining potential profile and MM geometry or domain, although not well characterized.

In conclusion, we demonstrate coherent manipulations of four individual spins in a linearly coupled QQD with the MM-ESR method.
The QQD is the largest multiple QD system ever used for coherent control of single electron spins.
From measurements of Rabi oscillations and ESR signals, we quantified variations of the $g$-factor and the MM induced stray field at the dot positions.
Our analysis hints at inhomogeneity in the spin-electric coupling, which may be due to the inhomogeneous dot potentials or MM stray field.
The results obtained here imply the gate-voltage-tunable addressability of four individual spins with the MM-ESR method, and therefore may pave the way towards the further scale-up of the spin qubit systems with QDs.


We thank the Microwave Research Group in Caltech for technical support.
This work was supported financially by Core Research for Evolutional Science and Technology (CREST), Japan Science and Technology Agency (JST) (JPMJCR15N2, JPMJCR1675) and the ImPACT Program of Council for Science, Technology and Innovation (Cabinet Office, Government of Japan).
T.I. acknowledges support from Materials Education program for the future leaders in Research, Industry, and Technology (MERIT).
T.O. T.N. and J.Y. acknowledge support from RIKEN Incentive Research Projects.
T.O. acknowledges support from Precursory Research for Embryonic Science and Technology (PRESTO) (JPMJPR16N3), JSPS KAKENHI grant numbers JP16H00817 and JP17H05187, Advanced Technology Institute Research Grant, the Murata Science Foundation Research Grant, Izumi Science and Technology Foundation Research Grant, TEPCO Memorial Foundation Research Grant, The Thermal and Electric Energy Technology Foundation Research Grant, The Telecommunications Advancement Foundation Research Grant, Futaba Electronics Memorial Foundation Research Grant and Foundation for Promotion of Material Science and Technology of Japan (MST) Foundation Research Grant.
S.T. acknowledges support by JSPS KAKENHI grant numbers JP26220710 and JP16H02204.


\end{document}